\documentclass{article}
\usepackage{spconf,amsmath,amssymb,graphicx}

\def\x{{\mathbf x}}

\title{COMPRESSED BEAMFORMING APPLIED TO B-MODE ULTRASOUND IMAGING}
\name{Noam Wagner$^{\star}$ \qquad Yonina C. Eldar$^{\star}$ \qquad Arie Feuer$^{\star}$ \qquad Zvi Friedman$^{\star \dagger}$}

\address{$^{\star}$ Technion-Israel Institute of Technology, Technion City, Haifa, Israel\\ $^{\dagger}$ GE Health-care, Haifa, Israel}

\begin{document}
\ninept
\maketitle
\begin{abstract}
Emerging sonography techniques often imply increasing in the number of transducer elements involved in the imaging process.   Consequently, larger amounts of data must be acquired and processed by the beamformer.  The significant growth in the amounts of data effects both machinery size and power consumption.  Within the classical sampling framework, state of the art systems reduce processing rates by exploiting the bandpass bandwidth of the detected signals.  It has been recently shown, that a much more significant sample-rate reduction may be obtained, by treating ultrasound signals within the Finite Rate of Innovation framework.  These ideas follow the spirit of Xampling, which combines classic methods from sampling theory with recent developments in Compressed Sensing.  Applying such low-rate sampling schemes to individual transducer elements, which detect energy reflected from biological tissues, is limited by the noisy nature of the signals.  This often results in erroneous parameter extraction, bringing forward the need to enhance the SNR of the low-rate samples.  In our work, we manage to achieve such SNR enhancement, by beamforming the sub-Nyquist samples obtained from multiple elements. We refer to this process as ``compressed beamforming".  Applying it to cardiac ultrasound data, we successfully image macroscopic perturbations, while achieving a nearly eight-fold reduction in sample-rate, compared to standard techniques.   

\end{abstract}
\section{Introduction}
\label{sec:01}  
Diagnostic sonography allows visualization of body tissues, by radiating them with acoustic energy pulses, transmitted from a transducer array.  Applying appropriate delays to the transmitting elements, in a process known as ``beamforming", the interfering wave pattern forms a narrow beam, along which most energy propagates. As it propagates in the tissue, echoes are scattered by density and propagation-velocity perturbations~\cite{Jensen01}.  These reflections are detected by the transducer array in a reception phase, where a second beamforming process is applied, aimed at localizing the scattering structures while improving Signal to Noise Ratio (SNR)~\cite{Szabo01}.  The reception beamforming process involves averaging the detected signals, after aligning them with time-varying delays.  Ultrasound systems perform this process digitally, requiring that the analog signals first be sampled. Confined to classic Nyquist-Shannon sampling theorem~\cite{Shannon01}, these systems must sample the signals at twice their baseband bandwidth, in order to avoid aliasing.  

The goal of our work is to produce ultrasound images from samples taken at rates far below these dictated by the classical framework. A first step in this direction was proposed in~\cite{Tur01}, where the signals detected in each element were assumed to satisfy the FRI property~\cite{Vetterli01}.  Applying the FRI framework,~\cite{Tur01, Gedalyahu01} derive schemes, which allow reconstruction of ultrasound signals from samples taken at sub-Nyquist rates.  These approaches follow the spirit of Xampling~\cite{Mishali01}, a term which refers to the combination of classical sampling theory with recent developments in  Compressed Sensing (CS).  However, applying such schemes to data reflected from biological tissues, the resulting signal parameters are often erroneous.  This is mainly due to the noisy nature of multiple reflections, also known as ``speckles"~\cite{Szabo01}, which are scattered by microscopic perturbations in the tissue.

In order to cope with the noisy nature of the signals detected in individual elements,~\cite{Wagner03} first introduces a multi-sensor Xampling scheme, which enhances the SNR of the low-rate samples.  This is achieved, by integrating traditional beamforming with FRI Xampling. The approach proposed in~\cite{Wagner03} is farther developed in~\cite{Wagner02}, resulting in a more straightforward sampling mechanism.  Both approaches, referred to by the term ``Compressed Beamforming", are derived and simulated for the special case of linear, B-mode scan.

In the current work, we generalize the method introduced in~\cite{Wagner03, Wagner02} to the case of polar scan.  Additionally, we propose reconstructing the sampled signal by applying CS techniques, rather than spectral analysis methods, traditionally used in the FRI framework~\cite{Vetterli01}.  Finally, we apply our approach to real cardiac ultrasound data, where we successfully image macroscopic perturbations, while achieving a nearly eight-fold reduction in sampling rate, compared to standard imaging techniques. 

This paper is organized as follows: in Section \ref{sec:02}, we outline the principles of beamforming in ultrasound imaging.  In Section \ref{sec:03} we discuss the FRI model in the context of ultrasound imaging, and its contribution to rate reduction when sampling signals at individual receivers.  Our multiple-receiver Xampling scheme, which integrates beamforming into the Xampling process, is presented in Section \ref{sec:04}. In Section \ref{sec:05} we introduce experimental results, obtained for cardiac ultrasound data.

\section{Beamforming in Ultrasound Imaging}
\label{sec:02}
We begin by outlining the beamforming process, lying at the heart of ultrasound imaging.  The process is carried out both in transmission and reception.  During transmission, modulated pulses are transmitted from all active array elements.  Applying appropriate delays to the transmitting elements, the propagating energy may be directed along a narrow beam, in any desired direction, the result of a constructive and destructive interference pattern.  The acoustic reciprocity theorem allows to perform a similar process during reception.  In this case, we regard the perturbations, which reflect the transmitted energy, as point transmitters.  Applying appropriate delays to the detected signals, the array may be focused to any coordinate, such that if, indeed, energy was scattered from that point, then the combined signals will undergo constructive interference.  Knowing that most energy propagates along a narrow beam, and knowing its propagation velocity, we may associate time instances to radiated points in the medium, thus optimizing the reception focal point at these instances.  The resulting process, in which time-varying delays are applied to the detected signals, prior to combining them for SNR enhancement, is referred to as ``Dynamic Focusing".  

\begin{figure}
\begin{minipage}[b]{1.00\linewidth}
 \centering
  \centerline{\includegraphics[width=5.0cm]{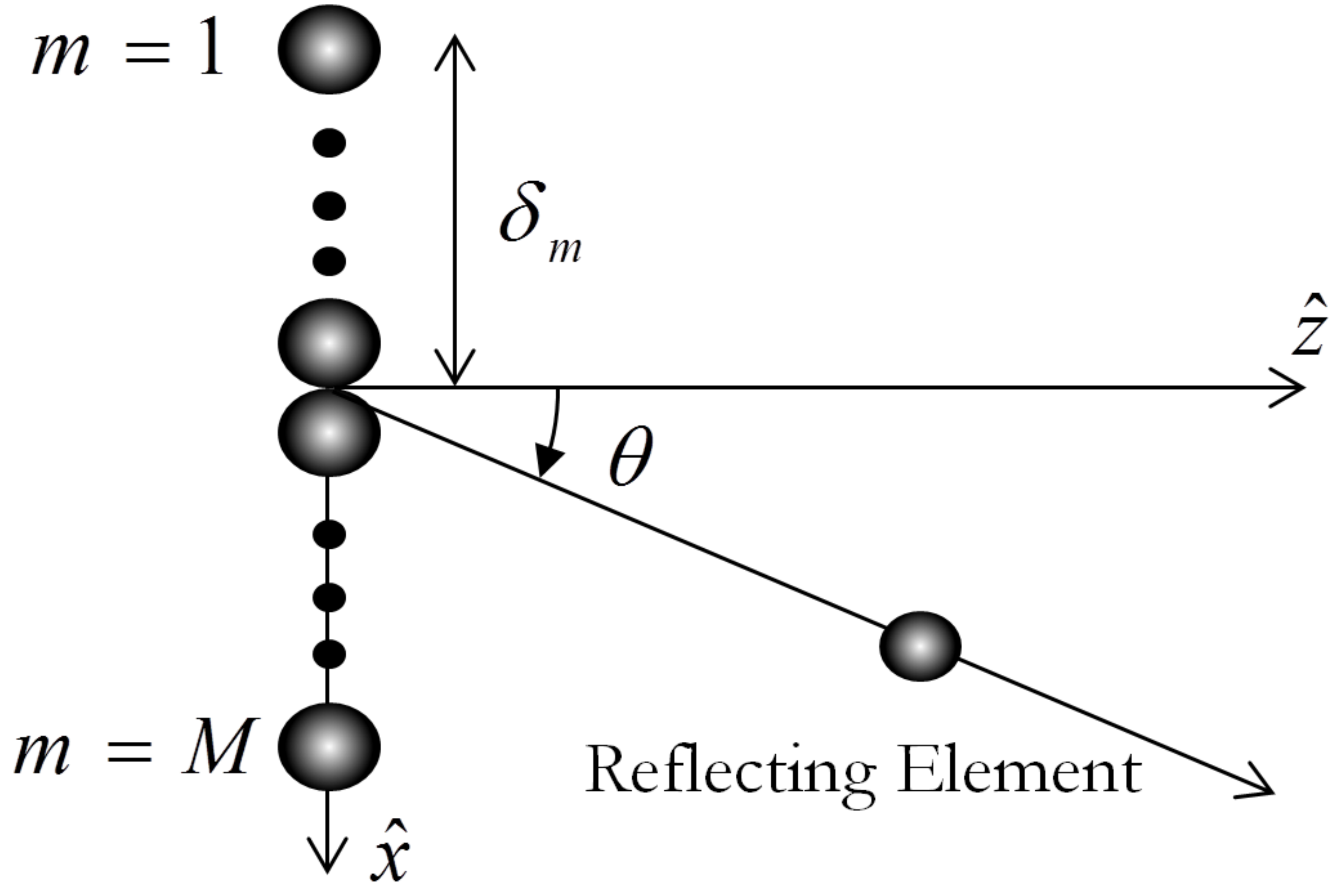}}
\end{minipage}
\caption{Imaging setup.}
\label{Fig:01}
\end{figure}

In order to derive an explicit expression for dynamic focusing, consider an array of $M$ receivers, aligned along the $\hat{x}$ axis.  We denote by $\delta_m$ the distance from the $m$'th receiver to the origin.  Assume a pulse of energy, transmitted at $t=0$ from the origin, at direction $\theta$.  The setup is depicted in Fig. \ref{Fig:01}.  At time $t$, the pulse, propagating at velocity $c$, crosses $\left(x,z\right)=\left(ct\cos\theta,ct\sin\theta\right)$.  A reflection originating from this coordinate will reach the $m$'th element at time:
\begin{equation}
\tau_m(t)=t+\sqrt{\left(t\cos\theta\right)^2+\left(\gamma_m-t\sin\theta\right)^2},
\end{equation}  
with $\gamma_m=\delta_m/c$.  
Denoting by $m_0$ the index of a reference receiver, satisfying $\delta_{m_0}=0$, it is readily seen that $\tau_{m_0}\left(t\right)=2t$.  Hence, in order to align the reflection detected in the $m$'th receiver with the one detected in  the reference receiver, we need to apply a delay to the detected signal, $\varphi_m\left(t\right)$, such that the resulting signal, ${\hat{\varphi}}_m\left(t;\theta\right)$, satisfies ${\hat{\varphi}}_m\left(2t;\theta\right)= \varphi_m\left(\tau_m\left(t\right)\right)$.  This implies a time-varying time delay, applied to $\varphi_m\left(t\right)$, resulting in the following distorted signal: 
\begin{align}
\label{E:02}
\begin{split}
{\hat{\varphi}}_m\left(t;\theta\right)&=\varphi_m\left(\frac{1}{2}\left(t+\sqrt{t^2-4\gamma_mt\sin\theta+4\gamma_m^2}\right)\right).  
\end{split}
\end{align}     
The aligned signals may be combined into the beamformed signal
\begin{equation}
\label{E:03}
\Phi\left(t;\theta\right)=\frac{1}{M}\sum_{m=1}^M{{\hat{\varphi}}_m\left(t;\theta\right)},
\end{equation}     
which exhibits enhanced SNR compared to $\left\{{\hat{\varphi}}_m\left(t;\theta\right)\right\}_{m=1}^M$.     

Although defined over continuous time, ultrasound systems perform the process formulated in \eqref{E:02}-\eqref{E:03} in the digital domain, requiring that the analog signals $\left\{\varphi_m\left(t\right)\right\}_{m=1}^M$ first be sampled.  Confined to classic Nyquist-Shannon sampling theorem, these systems sample the signals at twice their baseband bandwidth, in order to avoid aliasing.  The detected signals typically occupy only a portion of their baseband bandwidth.  Exploiting this fact, some state of the art systems manage to reduce the amount of samples transmitted from the front-end, by down-sampling the data, after demodulation and low-pass filtering.  However, since such operations are carried out digitally, the preliminary sampling-rate remains unchanged.  

\section{Sample Rate Reduction Using the FRI Model}
\label{sec:03}
Our goal is to produce an ultrasound image from samples taken far below the Nyquist rate.  A first step in this direction was proposed in \cite{Tur01}, where the signals detected in each element were assumed to satisfy the FRI property.  More specifically, they were approximated as $L$ replicas of a known-shape pulse, $h\left(t\right)$, caused by scattering of the transmitted pulse from $L$ reflectors, located along the narrow transmission beam.  Explicitly, the signal detected in the $m$'th receiver was modeled as
\begin{equation}
\label{E:04}
\varphi_m\left(t\right)=\sum_{l=1}^L{a_{l,m}h\left(t-t_{l,m}\right)},
\end{equation}
where $t_{l,m}$ denotes the time, in which the reflection, originating from the $l$'th reflector, arrives at the $m$'th receiver, and $a_{l,m}$ denotes the reflection amplitude.  Hence, the detected signal is completely defined by $2L$ parameters, $\left\{t_{l,m},a_{l,m}\right\}_{l=1}^L$. As shown in \cite{Vetterli01}, spectral analysis techniques (e.g. annihilating filter~\cite{Stoica01}) may be applied, in order to extract the signal parameters, from its projection onto a $2L$-dimensional subspace, corresponding to its Fourier series coefficients.  The sampling scheme is thus reduced to the problem of extracting $K$ frequency samples of the detected signal, where $K\geq2L$.  Denote the temporal support of the detected signal, $\varphi_m\left(t\right)$, by $\left[0,T\right)$.  Imaging to a nominal depth of $16 \mbox{cm}$, $T$ is approximately $210\mu\mbox{sec}$.  The signal's baseband bandwidth requires a nominal sampling rate of $f_s=16\mbox{Mhz}$, resulting in an overall number of $Tf_s=3360$ real-valued samples.  Still within classical sampling framework, assuming that the signal's passband bandwidth is only $4\mbox{MHz}$, the data sampled at Nyquist rate may be down-sampled to 1680 real-valued samples.   Thus, as long as the number of reflected pulses satisfies $L\ll1680$, substantial rate reduction may be achieved by an FRI Xampling scheme.       

The scheme proposed in \cite{Tur01} enables sub-Nyquist sampling of the signal detected in an individual array element.  Sampling signals reflected from phantom targets, it managed to achieve good signal reconstruction.   However, when applied to data reflected from biological tissues, the resulting signal parameters are often erroneous.  This is mainly the result of multiple reflections, scattered from microscopic perturbations, manifested as noise. In extreme scenarios, where the noise masks the FRI component, the extracted parameters will be meaningless, such that any attempt to cope with errors in the parametric space will turn out useless.  

The novelty of our approach regards the assumption, that the FRI scheme applied to an individual element may be equally applied to the beamformed signal defined in \eqref{E:03}, which obviously exhibits much better SNR.  Indeed, it may be shown that, if the signals $\left\{\varphi_m\left(t\right)\right\}_{m=1}^M$ obey the FRI model \eqref{E:04}, then $\Phi\left(t;\theta\right)$ is {\bf{approximately}} of the form: 
\begin{equation}
\label{E:05}
\Phi\left(t;\theta\right)=\sum_{l=1}^L{b_l h\left(t-t_{l}\right)},
\end{equation}
and may thus be regarded as FRI.  The parameters $\left\{t_l,b_l\right\}_{l=1}^L$ in \eqref{E:05} depend on $\theta$, yet we omit this dependency from our notation, for convenience.  An obvious problem is that, as stated in Section \ref{sec:02},  the signal $\Phi\left(t;\theta\right)$ does not exist in the analog domain. In the next section, we derive our scheme, which manages to estimate its necessary low-rate samples from low-rate samples of filtered versions of   $\left\{\varphi_m\left(t\right)\right\}_{m=1}^M$.    
      
\section{Compressed Beamforming Scheme}
\label{sec:04}
The FRI Xampling scheme applies spectral analysis techniques, in order to estimate the $2L$ unknown signal parameters, from a subset of its consecutive Fourier coefficients.  Denoting the cardinality of this subset by $K$, perfect reconstruction requires that $K\geq2L$.  

Denote by $\kappa$ the set of indices $\left\{k_j\right\}_{j=1}^K$, corresponding to the selected Fourier coefficients.  Having obtained these coefficients, other mechanisms may be applied, in order to reconstruct the signal.  In our  work, for instance, we  use CS methodology.  Since each of the signals $\left\{\varphi_m\left(t\right)\right\}_{m=1}^M$ is supported on $\left[0,T\right)$, then, by \eqref{E:02}-\eqref{E:03}, $\Phi\left(t;\theta\right)$ may only be constructed on a support  contained in $\left[0,T\right)$.  Considering its Fourier series, calculated with respect to this interval, let us explicitly write the $k_j$'th Fourier coefficient, $c_j$.  We will additionally assume, that the time delays, $\left\{t_l\right\}_{L=1}^L$, are quantized with $\Delta_s$ quantization step, such that $t_l=q_l\Delta_s$, $q_l\in\mathbb{Z}$:
\begin{align}
\label{E:06}
c_j&=\frac{1}{T}\int_{0}^{T}{\sum_{l=1}^L{b_l h\left(t-t_l\right)}e^{-j\frac{2\pi}{T}k_j t}dt} \notag \\
&=\frac{1}{T}\sum_{l=1}^L{b_l\int_{0}^{T}{h\left(t-q_l\Delta_s\right)}e^{-j\frac{2\pi}{T}k_j t}dt}\notag \\
&=\frac{1}{T}H\left(\frac{2\pi}{T}k_j\right)\sum_{l=1}^L{b_l e^{-j\frac{2\pi}{T}\Delta_s k_j q_l}}.
\end{align}
Here $H\left(\omega\right)$ denotes the CTFT of $h\left(t\right)$.  Let $N$ be the ratio $\lfloor T/\Delta_s\rfloor$.  Then \eqref{E:06} may be brought into the following matrix form: 
\begin{equation}
\label{E:07}
{\bf{c}}=\frac{1}{T}{\bf{HAx}},
\end{equation}
where ${\bf{H}}$ is the $K\times K$ diagonal matrix with $H\left(\frac{2\pi}{T}k_j\right)$ as its $j$'th element, and ${\bf{x}}$ is a length $N$ vector, whose $j$'th element equals $b_l$ for $j=q_l$, and $0$ otherwise.  Finally, ${\bf{A}}$ is a $K\times N$ matrix, formed by taking the set $\kappa$ of rows from a $N\times N$ FFT matrix.  Assuming that $L\ll N$, this is a CS problem, aimed at estimating the $L$-sparse vector ${\bf{x}}$ from $K$ samples, taken with a measurement matrix $\frac{1}{T}{\bf{HA}}$.  A common technique for recovering $\x$ is the Orthogonal Matching Pursuit (OMP) algorithm, which we use in our work.

We now return to the problem of obtaining the set $\kappa$ of $\Phi\left(t;\theta\right)$'s Fourier coefficients.  Let $c_j$ denote the $j$'th Fourier coefficient of $\Phi\left(t;\theta\right)$.  Then,
\begin{align}
\begin{split}
\label{E:09}
c_j&=\frac{1}{M}\sum_{m=1}^{M}{\frac{1}{T}\int_0^T{e^{-j\frac{2\pi}{T}k_j t}{\hat{\varphi}}_m\left(t;\theta\right)dt}}=\frac{1}{M}\sum_{m=1}^{M}{c_{j,m}}, 
\end{split}
\end{align}
where, from \eqref{E:02}:
\begin{align}
\begin{split}
\label{E:10}
c_{j,m}&=\\
&\frac{1}{T}\int_0^T{e^{-j\frac{2\pi}{T}k_j t}\varphi_m\left(\frac{1}{2}\left(t+\sqrt{t^2-4\gamma_mt\sin\theta+4\gamma_m^2}\right)\right)dt}.  
\end{split}
\end{align}
Applying a change of variable to the integral,we can write \eqref{E:10} as
\begin{align}\label{E:11}    
c_{j,m}&=\frac{1}{T}{\int_{0}^{T}g_{j,m}(t;\theta)\varphi_m\left(t\right)dt},
\end{align}
with
\begin{align}\label{E:12}    
\begin{split}
g_{j,m}(t;\theta)=&q_{j,m}(t;\theta)e^{-i\frac{2\pi}{T}k_j t},\\ 
q_{j,m}(t;\theta)=&I_{\left[|\gamma_m|,T_m\left(\theta\right)\right)}\left(t\right)\left(1+\frac{\gamma_m^2  \cos^2\theta}{\left(t-\gamma_m \sin\theta\right)^2}\right) \times \\
&\exp\left\{i \frac{2\pi}{T} k_j\frac{\gamma_m-t\sin\theta}{t-\gamma_m\sin\theta}\gamma_m \right\}.
\end{split}
\end{align}
Here $I_{\left[a,b\right)}\left(t\right)$ denotes an indicator function having the value 1 for $a\leq t<b$ and 0 otherwise, and $T_m\left(\theta \right)$ is the maximal time value which is mapped, under the temporal distortion defined in \eqref{E:02}, to the support of the beamformed signal.  

The process defined in \eqref{E:11}-\eqref{E:12} can be translated into a multi-channel Xampling scheme, in which each signal, $\varphi_m\left(t\right)$, is multiplied by a bank of kernels $\left\{g_{j,m}\left(t;\theta\right)\right\}_{j=1}^K$ defined by \eqref{E:12}, and integrated over $\left[0, T\right)$.  This results in a vector ${\bf{c_m}}=\left[\begin{array}{cccc}c_{1,m}&c_{2,m}&...&c_{K,m}\end{array}\right]^T$.  The vectors $\left\{{\bf{c_m}}\right\}_{m=1}^M$ are then averaged in ${\bf{c}}=\left[\begin{array}{cccc}c_{1}&c_{2}&...&c_{K}\end{array}\right]^T$, which has the desired improved SNR property, and provides a basis for extracting the $2L$ parameters which define $\Phi\left(t;\theta\right)$.  Nevertheless, the complexity of the analog kernels, together with their dependency on $\theta$, makes hardware implementation of such a scheme extremely complicated.  We thereby take an additional step, approximating the samples $\left\{c_{j,m}\right\}_{j=1}^K$ from low-rate samples of $\varphi_m\left(t\right)$, taken in a much more straightforward manner. 

First, we substitute $\varphi_m\left(t\right)$ of \eqref{E:11} by its Fourier series, calculated with respect to $\left[0,T\right)$.  Denoting the $n$'th Fourier coefficient by $\phi_m\left[n\right]$, we get: 
\begin{align}\label{E:13}    
\begin{split}
c_{j,m}&=\sum_{n}{\phi_m\left[n\right]\frac{1}{T}\int_0^T{q_{j,m}(t;\theta)e^{-i\frac{2\pi}{T}\left(k_j-n\right) t}dt}}\\
&=\sum_{n}{\phi_m\left[k_j-n\right]Q_{j,m;\theta}\left[n\right]},
\end{split}
\end{align}
where $Q_{j,m;\theta}\left[n\right]$ are the Fourier series coefficients of $q_{j,m}(t;\theta)$, also defined on $\left[0, T\right)$. Replacing the infinite summation of \eqref{E:13} by its finite approximation:
\begin{align}\label{E:14}
\begin{split}
{\hat{c}}_{j,m}=\sum_{n=N_1}^{N_2}{\phi_m[k_j-n]Q_{j,m;\theta}\left[n\right]},
\end{split}
\end{align}  
it can be shown that, for any $\epsilon>0$, and for any selection $\left(j,m;\theta\right)$, there exist $N_1\left(\epsilon, k_j, m; \theta \right)$ and $N_2\left(\epsilon, k_j, m;\theta \right)$ such that $|{\hat{c}}_{j,m}-c_{j,m}|<\epsilon$.  Therefore, with the finite sum \eqref{E:14}, we  may approximate $c_{j,m}$ to any accuracy.  Furthermore, setting an upper bound on the energy of $\varphi_m\left(t\right)$, $N_1$ and $N_2$ may be chosen off-line, subject to the decay properties of the sequence $\left\{Q_{j,m;\theta}\left[n\right]\right\}_{n=-\infty}^{\infty}$.  

\begin{figure}
\begin{minipage}[b]{1.00\linewidth}
 \centering
  \centerline{\includegraphics[width=8.5cm]{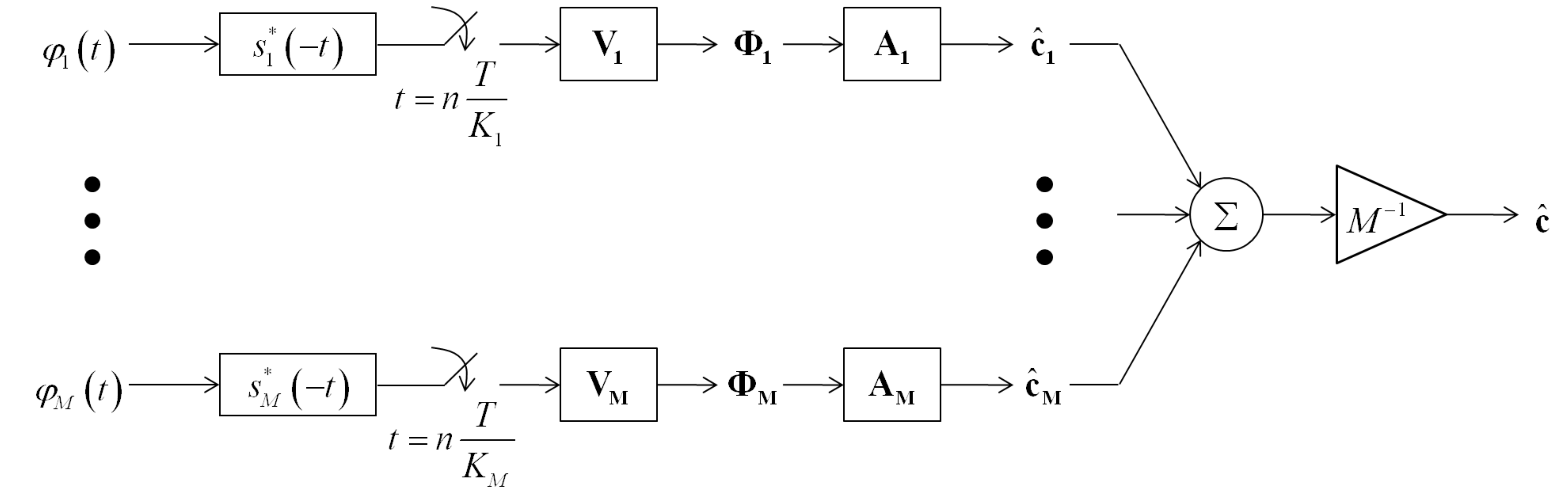}}
\end{minipage}
\caption{Xampling scheme utilizing Fourier samples of detected signals.}
\label{Fig:02}
\end{figure}

With this result we can, for each $\theta$ and receiver index $m$, find the minimal set of $\varphi_m\left(t\right)$'s Fourier series coefficients, from which we can simultaneously approximate the set $\left\{c_{j,m}\right\}_{j=1}^K$.  Denoting the set of indices corresponding to these coefficients by $\kappa_m$, arrange the set $\left\{\phi_m\left[k\right]\right\}_{k\in\kappa_m}$ in a length $K_m$ vector ${\bf{\Phi_m}}$.  Using \eqref{E:14}, we may define a linear transformation ${\bf{A_{m}}}\left(\theta\right)$, such that:
\begin{align}\label{E:15}
\begin{split}
{\bf{{\hat{c}}_m}}={\bf{A_{m}}}\left(\theta\right){\bf{\Phi_m}},
\end{split}
\end{align}  
where ${\bf{\hat{c}_m}}$ is the length $K$ vector containing $\hat{c}_{j,m}$ as its $j$'th element.  Instead of extracting only $K$ samples from each element using the complicated analog kernels, we now extract a larger number of samples, $K_m$.  As we show in our simulation, good approximation is obtained with just a small sampling overhead.  Our final Xampling scheme is illustrated in Fig. \ref{Fig:02}.  Note the simple mechanism used for obtaining the Fourier coefficients in each individual element:  a linear transformation, ${\bf{V_m}}$, is applied to point-wise samples of the signal, taken at a sub-Nyquist rate, after filtering it with an appropriate kernel, $s_m^*\left(-t\right)$, such as the Sum of Sincs ~\cite{Tur01}. 
\begin{figure*}
\begin{minipage}{.33\linewidth}
\centering
\centerline{\includegraphics[width=5.0cm]{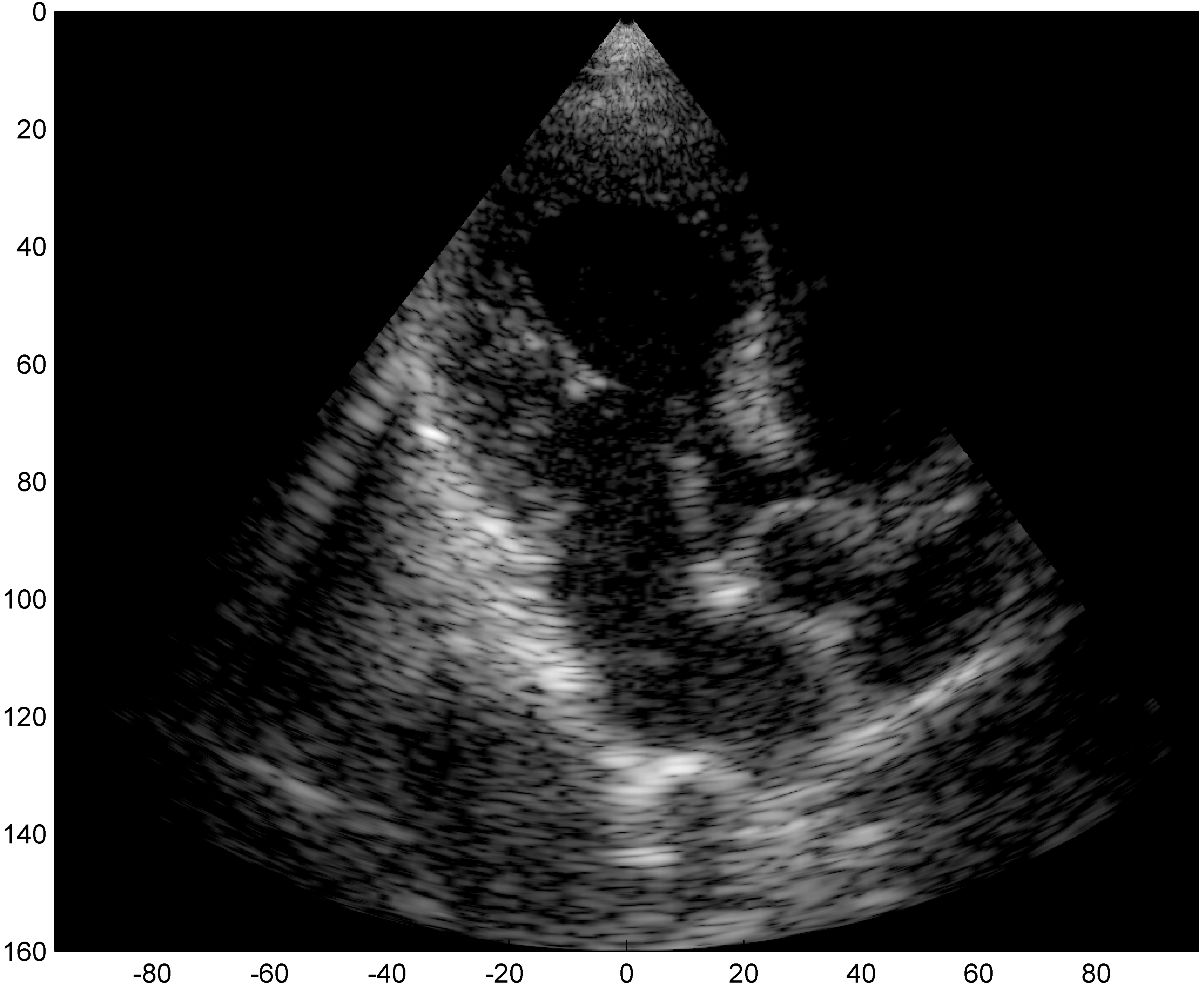}}
\centerline{(a)}
\end{minipage}
\begin{minipage}{.33\linewidth}
\centering
\centerline{\includegraphics[width=5.0cm]{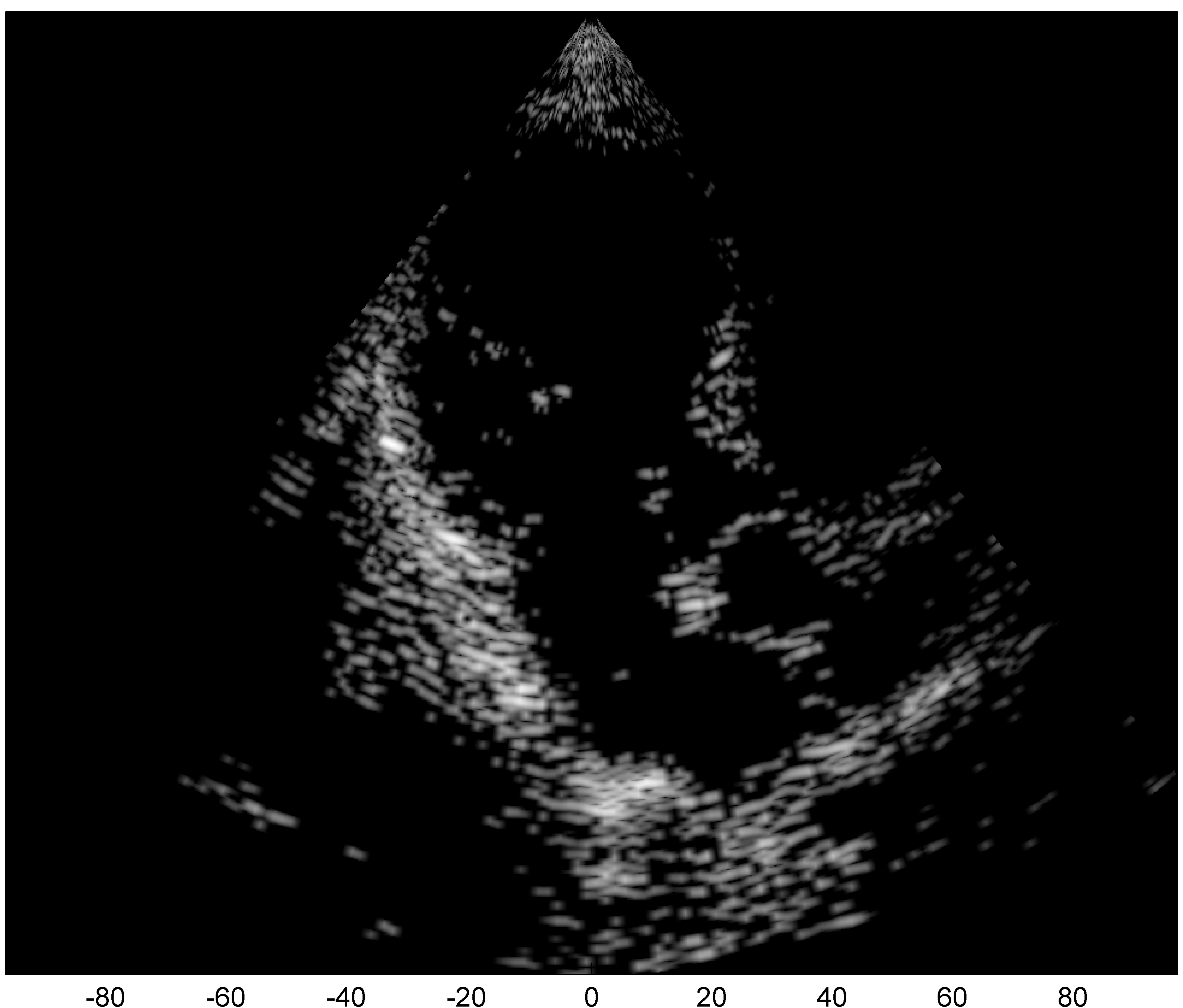}}
\centerline{(b)}
\end{minipage}
\begin{minipage}{.34\linewidth}
\centering
\centerline{\includegraphics[width=5.0cm]{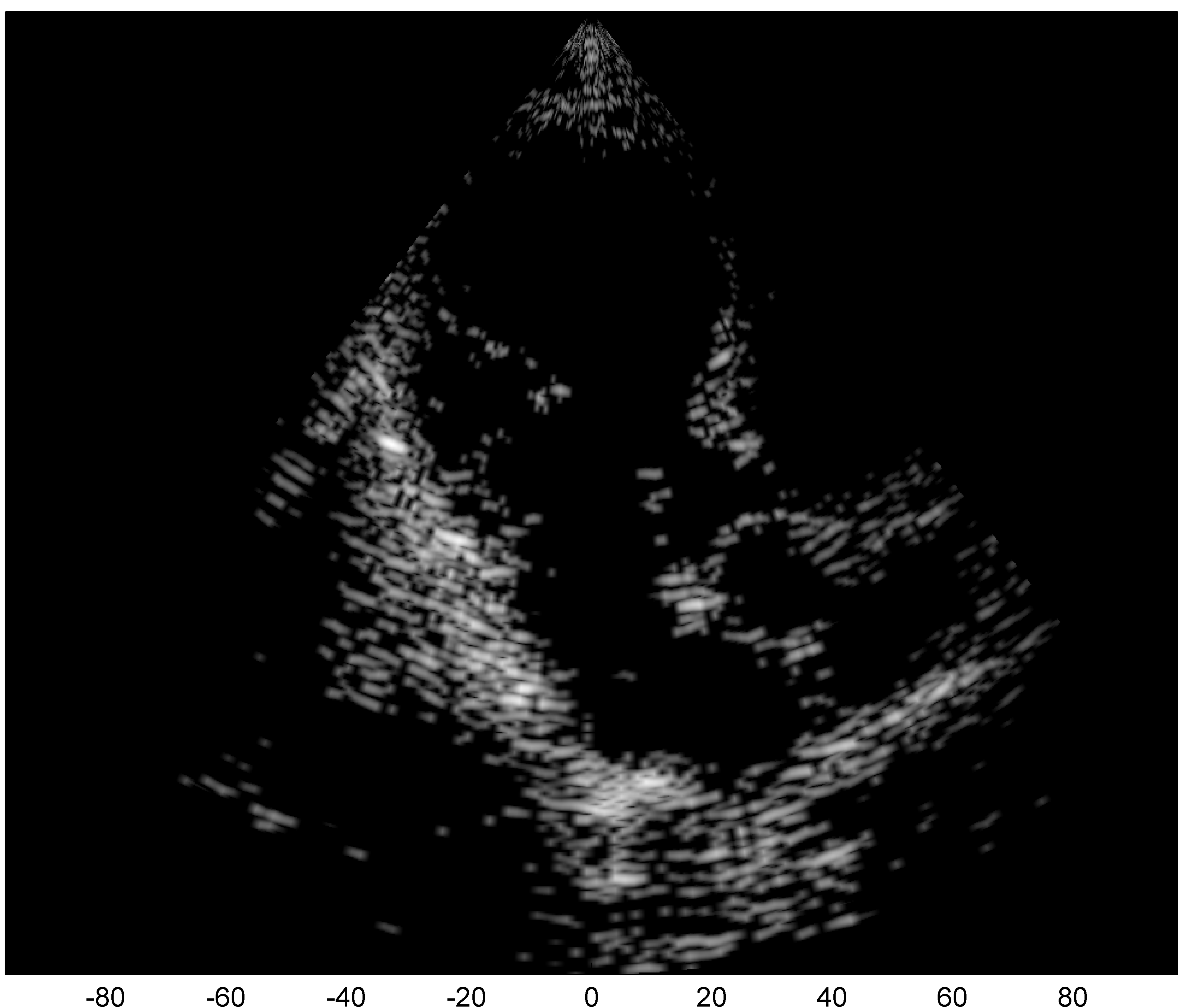}}
\centerline{(c)}
\end{minipage}
\caption{Cardiac images generated by Xampling and using traditional methods. (a) standard beamforming applied to data sampled at Nyquist rate.  (b) applying the non-approximated Xampling scheme, defined in \eqref{E:11}-\eqref{E:12}.  (c) applying the final Xampling scheme of Fig.~\ref{Fig:02}.}
\label{Fig:03}
\end{figure*}

\section{Simulation On Cardiac Ultrasound Data}	
\label{sec:05}
We examine the result of applying our scheme of Fig. \ref{Fig:02} to raw RF data, acquired and stored for cardiac images of a healthy consenting volunteer.  The acquisition was performed using a breadboard ultrasonic scanner of 64 acquisition channels.  The transducer employed was a 64-element phased array probe, with $2.5\mbox{MHz}$ central frequency,  operating in second harmonic imaging mode: 3 half cycle pulses are transmitted at $1.7\mbox{MHz}$, resulting in a signal characterized by a rather narrow bandpass bandwidth, centered at $1.7\mbox{MHz}$.  The corresponding second harmonic signal, centered at $3.4\mbox{MHz}$, is then acquired.  The signal detected in each acquisition channel is amplified and digitized at a sampling-rate of $50\mbox{MHz}$.   Data from all acquisition channels were acquired along 120 beams, forming a $60^{\circ}$ sector, where imaging to a depth of $z=16\mbox{cm}$, we have $T=207{\mu\mbox{sec}}$.  The results are illustrated in Fig. \ref{Fig:03}. 

The first image (a) was generated using the standard technique, applying beamforming to data first sampled at the Nyquist rate, and then down-sampled, exploiting its limited essential bandwidth.  For a single image line, sampling at $50 {\mbox{MHz}}$, we acquire $10389$ real-valued samples from each element, which are then down-sampled, to $1662$ real-valued samples, used for beamforming.  The resulting image is used as reference, which we aim at reproducing with our Xampling scheme.   We begin by applying our scheme without approximation, simulating the modulation with the analog kernels defined in \eqref{E:12}.  Assuming $L=25$ reflectors, and using two-fold oversampling, $\kappa$ comprises $K=100$ indices.  Since each sample is complex, we get an eight-fold reduction in sample-rate.  The resulting image (b) well depicts strong perturbations observed in (a).  Isolated reflectors at the proximity of the array ($z\approx 6{\mbox{cm}}$) remain well in focus.  Applying our approximated scheme, for every $k_j\in\kappa$, $1\leq m \leq M$ and $\theta$, we choose $N_1$ and $N_2$ of \eqref{E:14} such that $\sum_{n=N_1}^{N_2}{\left|Q_{j,m;\theta}\left(\frac{2\pi}{T}n\right)\right|^2}\approx 0.95\sum_{n=-\infty}^{\infty}{\left|Q_{j,m;\theta}\left(\frac{2\pi}{T}n\right)\right|^2}$.  At the current stage, this choice, and consequently the choice of $\left\{\kappa_m\right\}_{m=1}^M$ and $\left\{{\bf{A_m}}\right\}_{m=1}^M$, are performed off-line, based on the imaging setup.  This results in an average number of $116$ complex samples per receiving element, required for constructing each image line (with at most $133$ samples required for some combinations of $\theta$ and element indices).  We arrive at a seven-fold reduction in sample rate, and the resulting image (c) appears very similar to (b).  
\bibliographystyle{IEEEbib}

\end{document}